\documentclass[reprint,amsmath,twocolumn,amssymb, nobibnotes, aps, pre, superscriptaddress]{revtex4-1}

\usepackage{graphicx}
\usepackage{dcolumn}
\usepackage{bm}
\usepackage{color}
\usepackage{tikz}
\usetikzlibrary{shapes}
\usetikzlibrary{backgrounds}
\usetikzlibrary{plotmarks}

\begin{document}
\def \beq{\begin{equation}}
\def \eeq{\end{equation}}
\def \bea{\begin{eqnarray}}
\def \eea{\end{eqnarray}}
\def \bem{\begin{displaymath}}
\def \eem{\end{displaymath}}
\def \P{\Psi}
\def \Pd{|\Psi(\boldsymbol{r})|}
\def \Pds{|\Psi^{\ast}(\boldsymbol{r})|}
\def \Po{\overline{\Psi}}
\def \bs{\boldsymbol}
\def \bl{\bar{\boldsymbol{l}}}

\title{{{Solitons, the Korteweg-de Vries equation with step boundary values 
and pseudo-embedded eigenvalues}}}
\author{M.J. Ablowitz}
\affiliation{Department of Applied Mathematics, University of Colorado, Boulder, Colorado 80309}
\author{X-D. Luo}
\affiliation{Department of Mathematics, State University of New York at Buffalo, Buffalo, New York 14260-2900}
\author{J.T. Cole}
\affiliation{Department of Applied Mathematics, University of Colorado, Boulder, Colorado 80309}

\begin{abstract}
The Korteweg-deVries (KdV) equation with step
boundary conditions is considered, with an emphasis on soliton dynamics. When one or more initial solitons
are of sufficient size they can propagate through the step; 
in this case the phase shift is calculated via the inverse scattering transform. On the other hand, when the amplitude is 
too small they become trapped.  In the trapped case the transmission coefficient of the associated associated linear Schr\"odinger equation can become  large at a point exponentially close to the continuous spectrum. This point is referred to as a {\it pseudo-embedded eigenvalue}. Employing the inverse problem it is shown that 
the continuous spectrum associated with a branch cut in the neighborhood of the pseudo-embedded eigenvalue plays the role of discrete spectra, which in turn leads to a trapped soliton in the KdV equation.

\end{abstract}



\maketitle

\section{Introduction}
The realization that  solitary waves are important dynamical entities goes back to {seminal} observations and experiments of shallow water waves by Russell \cite{Russ1844}. Motivated by these observations Korteweg and deVries  \cite{KdV1895} discovered an important equation in shallow water waves. 
 Indeed the KdV equation is a universal nonlinear system which arises whenever there is  a balance of weak
dispersion and quadratic nonlinearity \cite{MJA2011}.

We write {KdV's} equation in the following {normalized} form
\begin{equation}
\label{E:KdV}
u_{t}+6uu_{x}+u_{xxx}=0 \; .
\end{equation}
In the context of water waves $t$ is time, $x$ is a 
spatial coordinate, and $u$ is the fluid velocity.
In 1965 Zabusky and Kruskal \cite{ZabKr1965} found that 
solitary waves of the KdV equation had special interaction properties; {namely} any two of them interacted elastically. They termed these waves {{\it solitons}}.
{Soon afterward,} Gardner, Greene, Kruskal and Miura (GGKM) \cite{GGKM1967} associated 
{eq.~(\ref{E:KdV})} with two {\it linear} equations.
One of these equations is the celebrated time{-}independent  Schr\"odinger equation of quantum mechanics
\begin{equation}
\label{E:Schrod}
v_{xx}+ (u(x,t)+k^2)v=0 ,
\end{equation}
where $u(x,t)$ is a real 
{potential,} $k^2$ is the wave energy; and $t$ is {treated as a parameter}. 

For rapidly decaying boundary conditions (BCs) GGKM showed how direct and inverse scattering of (\ref{E:Schrod}) could be used to linearize {and solve} the KdV equation. {In particular, they} 
showed that solitons were related to time{-}independent eigenvalues/bound states of eq.~(\ref{E:Schrod}) and obtained pure soliton solutions explicitly. The linearization is in terms of a Gel'fand-Levitan-{Marchenko} (GLM) integral equation which provides the inverse scattering/reconstruction of the 
solution 
$u(x,t)$ to eq.~(\ref{E:KdV}).  
{This method of solution is now called the Inverse Scattering Transform (IST) and considerable research using these techniques has ensued and continues today cf. \cite{Ablowitz2,NOVIKOV,MJA2011}}.
 {The analytical underpinnings of the 
 {direct/inverse scattering problems} associated with eq.~(\ref{E:Schrod}) {for} decaying {boundary} data can be found in \cite{Faddeev67,Deift79,Marchenko2011}. }

Most research associated with the KdV equation has been {posed on spatial domains with either decaying or periodic boundary values.}
{There is, however,}
an important related problem, which is 
eq.~(\ref{E:KdV}) subject to step 
 {BCs}
\begin{equation}
\label{E:boundary conditions}
\lim_{x\rightarrow -\infty}u=0, \ \ \ \lim_{x\rightarrow +\infty}u=\pm c^{2},
\end{equation}
where $c>0$ is constant and $u$ goes to these limits sufficiently 
 fast; we require that 
\begin{equation}
\int_{-\infty}^{\infty} |u(x,t)\mp c^{2} H(x)|(1+x^{2})dx < \infty,
\end{equation}
where $H(x)$ is the Heaviside function.
We refer to the increasing boundary condition $+c^2$ case as ``step up'' and  the decreasing boundary condition $-c^2$ case as ``step down''. {Since the KdV equation is  Galilean invariant, it suffices to consider $u \rightarrow 0$ as $x \rightarrow -\infty$}; i.e. any
{nonzero boundary condition $u \rightarrow u_0 \not=0$ as $x \rightarrow - \infty$ can be made zero through the transformation $u(x,t) = u_0 + \tilde{u}(x - 6 u_0 t, t).$}

The step  problem has 
been studied by a number of authors. With step down boundary values, the basic direct/inverse scattering theory of the Schr\"odinger equation was developed over 50 years ago by Buslaev and Fomin \cite{BF62}. Their results were later used to discuss the asymptotic behavior of certain solutions to 
the KdV equation in \cite{Hruslov76}. Subsequently the problem was studied by a number of authors in \cite{Cohen79,Cohen84,CohenKaep85} and later by \cite{Atko1999,Toesch2015} with the main aim of developing a rigorous understanding of the direct/inverse scattering of this problem.

 In terms of the KdV wave dynamics, pure step down data for $t>0$ leads to the development of collisionless or dispersive shock waves (DSWs) \cite{GurPit73}, whereas pure step up data for $t>0$ leads to the development of a {{linear}} ramp from $u=0$ to $u=c^2$ with small associated oscillations \cite{MJAJCXL2017}.  The asymptotic development  of dispersive shock waves due to multi-step {down} initial data was considered in  \cite{Baldwin2009,Baldwin2013}. 

Unlike previous research on the step BC problem,  we focus on the dynamical 
situation when solitons/${\rm sech^2}$ profiles in addition to a step are initially given.
To be concrete we use as initial data a delta function, a box, or a soliton/sech$^2$ profile located well to the left of a step/Heaviside function.
The step problem is different from the decaying to zero problem since we do not have `pure solitons', i.e. along with solitons there is always additional continuous spectrum (no reflectionless potential).  

In the context of equation (\ref{E:KdV}) there are two types of pulses associated with localized initial data positioned 
well to the left of a step. 
The first case is that of a pulse with a large enough amplitude that allows it to pass all the way thorough the step. 
The amplitude is related to the discrete spectra/eigenvalues. These are `proper' eigenvalues and correspond to 
zeros of the inverse of the transmission coefficient, i.e. poles of the transmission coefficient. Proper eigenvalues are associated with standard solitons that, as mentioned above, propagate through through the step with only a suitable Galilean shift (velocity increase) and phase shift. From the IST we derive the phase shift of the soliton as it passes  through the rarefaction ramp that evolves from an initial step up 
or the DSW in the step down case, 
with or without other solitons. 
The phase shift formulae are similar to those derived for the decaying problem with continuous spectrum \cite{AblKod82}. As an example, we find the phase shift for a single 
soliton passing through a step. 
 The J-soliton phase shift can be calculated by similar methods (see Appendix \ref{appendC}).

However, if the initial localized 
profile is not large enough we find that the inverse of the transmission coefficient has no zeros i.e. no proper eigenvalues. What we do find in this case are spectral values that are {\it exponentially} close to continuous spectra.
We term such a point 
a {\it pseudo-embedded eigenvalue}. Such pseudo-embedded eigenvalues are associated with soliton-like pulses which propagate as though they were true solitons for a while, but eventually become trapped inside the rarefaction ramp for the step up initial condition, 
 or the DSW in the step down case; hence they have interesting physical manifestations. In this paper we discuss the step up case; the step down case is similar.

Below we show that in the trapped case the `branch cut' term associated with the continuous spectrum in the the inverse scattering problem leads to a contribution that plays the role of discrete spectra. Said differently, the pseudo-embedded eigenvalue leads to a dominant contribution from the branch cut associated with the continuous spectrum that has exactly the form as 
 discrete spectra from a proper eigenvalue. This term gives rise to a pulse which travels uniformly like a soliton until it encounters the rarefaction ramp/DSW where it eventually becomes trapped. The pseudo-embedded eigenvalue provides a spectral interpretation of the trapped soliton.

We point out that in recent experiments \cite{Hoefer2016} solitons have been transmitted through rarefaction waves (step up) and dispersive shock waves (step down), moreover, it has been shown that small amplitude solitons can become trapped in rarefaction ramp/DSW that develops from step initial data.

In this paper we concentrate on step up BCs; i.e. eq.~(\ref{E:boundary conditions})  with the positive sign.
The analogous theory can be developed for step down BCs; see also additional remarks in the conclusion of this article.
These studies were motivated by lectures by M. Hoefer discussing analytical/experimental research summarized in \cite{Hoefer2016}.

\section{Scattering/Inverse scattering theory and KdV solitons}


The  KdV eq.~(\ref{E:KdV}) is the 
compatibility condition (Lax pair) for 
the following two linear equations {
\begin{equation}
\label{E:lax 1}
v_{xx}+\left(u(x,t)+k^{2}\right)v=0,
\end{equation}
}\begin{equation}
\label{E:time evolution}
v_{t}=\left(u_{x}(x,t)+\gamma\right)v+\left(4k^{2}-2u(x,t)\right)v_{x},
\end{equation}
where $k$ is the spectral parameter, $\gamma$ is a constant; the potential $u(x,t)$ satisfies the BCs (\ref{E:boundary conditions}) with a plus sign.

\subsection{{Eigenfunctions}}
Eigenfunctions of (\ref{E:lax 1}) are defined by the following {BCs} {
\begin{equation}
\label{E:asymptotic 1}
\phi(x,k)\sim e^{-ikx},~ \bar{\phi}(x,k)\sim e^{ikx}   \text{~as~} x \to -\infty , 
\end{equation}
\begin{equation}
\label{E:asymptotic 2}
\psi(x,\lambda)\sim e^{i\lambda x}, ~~ \overline{\psi}(x,\lambda)\sim e^{-i\lambda x}   \text{~as~} x\rightarrow  + \infty , 
\end{equation}
} where $k, \lambda$ are real and  
\begin{equation}
\label{lambda}
\lambda(k) = (k^{2}+c^{2})^{1/2} . 
\end{equation}
 {We take the branch cut of $\lambda(k)$ to be $k\in [-ic, ic]$, and the branch cut of $k(\lambda)$ to be $\lambda \in [-c, c]$; then $\Im k\geq 0$ when $\Im \lambda \geq 0$ and $\Im k\leq 0$ when $\Im \lambda \leq 0$.}
 From the governing integral equations, the eigenfunctions $\phi, \psi$ can be analytically continued into the upper half plane (UHP) of $k,\lambda$, while 
$\overline{\phi}, \overline{\psi}$ can be analytically continued into the corresponding lower half plane (LHP).
From eq.~(\ref{E:lax 1}) and the BCs (\ref{E:asymptotic 1})-(\ref{E:asymptotic 2}) we see that  the eigenfunctions are related by: 
\begin{align}
\label{E:symmetry}
&\phi(x,k)=\overline{\phi}(x,-k)=
\phi^*(x,k), \\
&\psi(x,\lambda)=\overline{\psi}(x,-\lambda) =
\psi^*(x,\lambda) ,
\end{align}
for $\lambda, k$ real and where asterisk represents complex conjugate. 
When $k=i\kappa, \kappa \in [-c, c]$
\begin{equation}
\phi(x,k)=\phi^*(x,k).
\end{equation}

\subsection{Scattering data}
The two eigenfunctions $\psi(x,\lambda),\overline{\psi}(x,\lambda)$  
are linearly independent for $k \neq 0$. Hence, we can write $\phi(x,k), \overline{\phi}(x,k) $ as a linear combinations of $\psi(x,\lambda)$ and $\overline{\psi}(x,\lambda)$. Thus, we have the relations, formulated on the left, termed the left 
scattering problem
\begin{equation}
\label{E:linear combination 1}
\phi(x,k)=a(k)\overline{\psi}(x,\lambda)+b(k)\psi(x,\lambda),
\end{equation}
\begin{equation}
\label{E:linear combination 2}
\overline{\phi}(x,k)=\overline{a}(k)\psi(x,\lambda)+\overline{b}(k)\overline{\psi}(x,\lambda) ,
\end{equation}
where $k$ and $\lambda$ are real.
The scattering data is given by
\begin{equation}
\label{E: scattering data}
a(k)=\frac{1}{2i\lambda}W(\phi, \psi), \ \ \ b(k)=\frac{1}{2i\lambda}W(\overline{\psi},\phi),
\end{equation}
where $W(u,v)=uv_{x}-vu_{x}$ is the Wronskian. We remark that from the relation $\lambda^2=k^2+c^2$ the scattering data $a,b$ can be written in terms of either $k$ or $\lambda$; i.e. $a=a(k)$ or $a=a(\lambda)$.
Similar relations hold for $\overline{a}(k),\overline{b}(k)$, and 
we can show
$\overline{b}(k)=b^*(k)$, $\overline{a}(k)=a^*(k)$ for $k$ real and $a(-\lambda)=b(\lambda)$ for real $\lambda$ such that $|\lambda|\leq c$. 
We note that $k=i\kappa, |\kappa| \leq c$  corresponds to real   $\lambda$, $|\lambda|\leq c$. This forms part of the continuous spectrum and plays an important role below.

For the left scattering problem (\ref{E:linear combination 1}) the usual transmission and reflection coefficients of quantum mechanics are, respectively,
\begin{equation}
\label{ReflTranCoef}
 \tau(k)=\frac{1}{a(k)}, ~~\rho(k)=\frac{b(k)}{a(k)}.
 \end{equation} 
 These correspond to a unit wave denoted by $e^{-ikx}$ propagating into the potential $u(x)$ from $x=-\infty$.

In the decaying problem, a soliton solution of the KdV equation is given by 
\begin{equation}
\label{soliton_soln}
u(x,t) = 2\kappa^2~\text{sech}^2 \left[ \kappa(x-4\kappa^2t-x_0) \right], ~ x_0 \in \mathbb{R} .
\end{equation}
 Solitons  are associated with zeros of $a(k)$: $k=i\kappa, \kappa>0$ such that $a(k) = 0$.  We also call $k = i \kappa$ an eigenvalue; it is related to a bound state of the Schr\"odinger eq.~(\ref{E:lax 1}). 

\begin{figure} [ht]
\includegraphics[scale=.39]{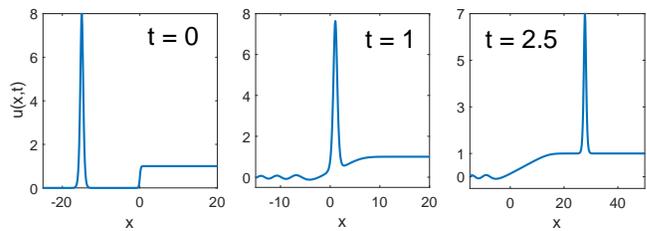}
\caption{Tunneling soliton example for $\kappa_1 = 2, c = 1, x_0 = -15$.}
\label{tunnel_soliton}
\end{figure}

The situation is quite different in the step problem. Consider a soliton initially positioned far to the left ($x_0 \ll -1 $) of a localized step centered at $x=0$. 
A soliton with amplitude parameter: $\kappa_1 > 0$, 
suggests it 
has a corresponding eigenvalue: $k=i\kappa_1$ with $\lambda_1$ given by
\begin{equation}
\label{soliton_eig}
 \lambda_1 = \sqrt{c^2- \kappa_1^2}  ~~ , ~~  \Im \lambda_1 > 0 .
\end{equation}
 When the `incoming' soliton has sufficient size, i.e. the corresponding eigenvalue $k=k_1 = i\kappa_1$ satisfies $\kappa_1>c$, 
 and $a(k_1) = 0, \lambda=\lambda_1 = i\eta_1, \eta_1>0$, then this is an instance of a proper eigenvalue and corresponds to a soliton tunneling through the rarefaction ramp (as an example, see Fig.~\ref{tunnel_soliton}). 
 An analogous equation was obtained via Whitham theory  in the weakly dispersive regime and was termed the `transmission condition'  \cite{Hoefer2016}.
 As in the decaying problem, the eigenvalue corresponds to a bound state; i.e the eigenmodes in eq.~(\ref{E:lax 1}) which are square integrable.  In this case we find the phase shift of the soliton.
 
\begin{figure} [ht]
\includegraphics[scale=.35]{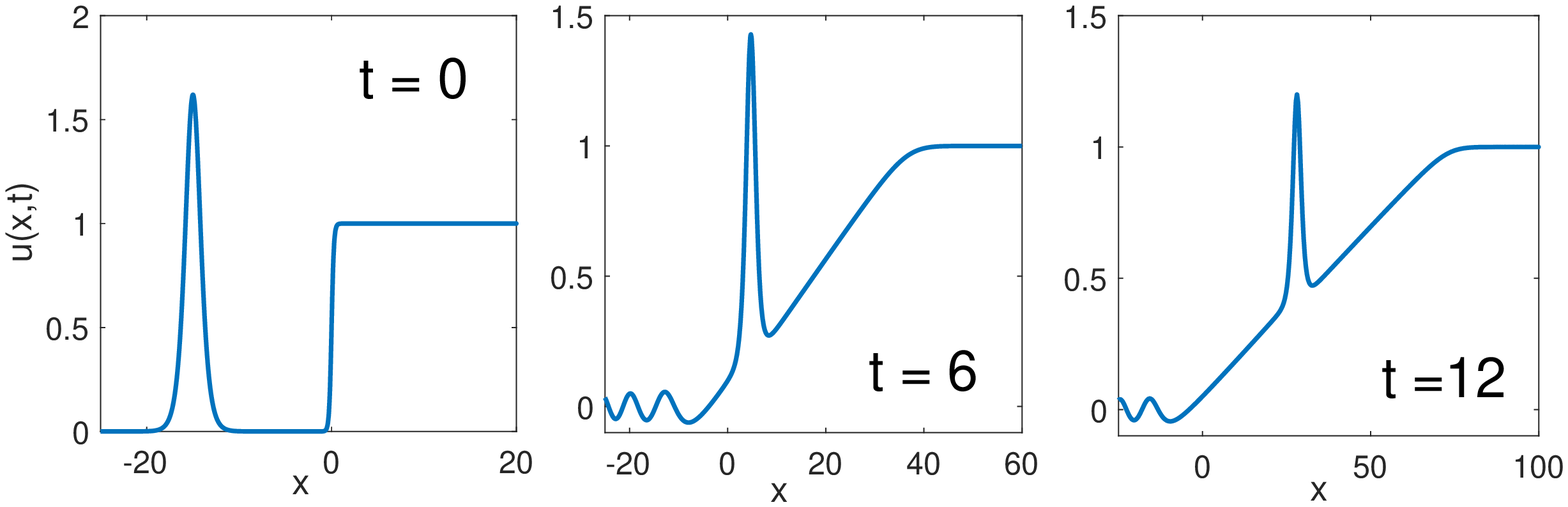}
\caption{Trapped soliton example for $\kappa_0 = 0.9, c = 1, x_0 = -15$.}
\label{trap_soliton}
\end{figure}

If, however, the initial soliton/${\rm sech}^2(x)$ profile is not 
large enough, then the `soliton' becomes trapped \cite{Hoefer2016,MJAJCXL2017}.  We term this a trapped soliton since it does look and travel like the ${\rm sech}^2$ solution in (\ref{soliton_soln}) to the left of the step. 
In fact we find this mode
eventually becomes trapped in the rarefaction ramp that evolves from the step (see Fig.~\ref{trap_soliton}), never reaching the top of the ramp. This is unlike a normal, or proper, soliton.
 Below we show that these trapped solitons correspond to {\it pseudo-embedded} eigenvalues 
 which are points that are exponentially close to the continuous spectrum. 
  
For an initial soliton/sech$^2$ profile, delta, or box function
we find the spectral coefficient $a(k)$ 
takes the form 
\begin{equation}
\label{a(k)}
a(k) = a_1(k)(k-i\kappa_0)+\epsilon a_2(k), ~ \epsilon=e^{2\kappa_0 x_0} \ll 1.
\end{equation}
with $0<\kappa_0 <c$ and $a_2(i \kappa_0) \neq 0;$ where $x_0$ is the initial position of a soliton or delta function or box. The calculations leading to equation (\ref{a(k)}) are discussed in Appendix \ref{appendA}. Since we take $x_0 \ll -1, ~ \epsilon$ is exponentially small. Although $a(k)$ is not found to be exactly zero {\it anywhere} in the upper half plane, it does possesses values of $k$, $ 0\leq \Im k \leq c$, that are exponentially close to 
the imaginary $k$ axis and thus have properties analogous to those of discrete spectra/eigenvalues. The value $k_0 \approx i \kappa_0$ provides the spectral meaning of a trapped soliton. 

 In Sec.~\ref{IST via GLM and Soliton Phase Shifts} we discuss the inverse scattering problem in the case of proper eigenvalues and proper solitons.
In Sec.~\ref{pseudo_emebed_sec} we show how the form of $a(k)$ given in (\ref{a(k)}) leads to a spectral contribution that plays the role of a discrete eigenvalue even though, by assumption, there are no proper eigenvalues/discrete spectra in the GLM equation. 
The scattering data for three prototypical examples (delta function, box function, soliton) are given in Appendix \ref{appendA}. 

\section{IST via GLM and Soliton Phase Shifts}
\label{IST via GLM and Soliton Phase Shifts}
The left  scattering problem in eq.~(\ref{E:linear combination 1}) can be transformed into a  GLM equation from $x ~\text{to}~ \infty$. To do this we assume  $\psi$ has the following triangular form
\begin{equation}
\label{E:integral equation 1}
\psi(x,\lambda;t)=e^{i\lambda x}+\int_{x}^{\infty}G(x,s;t)e^{i \lambda s}ds .
\end{equation}
Substituting the above representation into eq.~(\ref{E:linear combination 1}), using (\ref{E:symmetry}), then dividing by
$a(k)$, integrating over 
$\mathbb{R}$ with $\int d\lambda \exp( i\lambda(y-x)) / (2 \pi)$ for $y>x$ and carrying out requisite calculations yields the following GLM equation,
\begin{equation}
\label{E:GLM+}
G(x,y;t)+\Omega(x+y;t)+\int_{x}^{\infty}\Omega(y+s;t)G(x,s;t)ds=0, 
\end{equation}
where the kernel is given by
\begin{equation}
\label{KernelGLM+}
\Omega(z;t)=\frac{1}{2\pi}\int_{-\infty}^{\infty}\rho(\lambda;t)e^{i\lambda z}d\lambda
-\sum_{j=1}^{J}c_{j}e^{i \lambda_{j}z} ,
\end{equation}
\and
\[ \rho(\lambda;t)= \frac{b(\lambda,t)}{a(\lambda,t)},~ c_{j}(t)=\frac{i b(\lambda_{j}; t)}{\partial_{\lambda}a(\lambda_{j}; t)},~~\Im \lambda_j>0 \]
with $\lambda_{j}, \Im \lambda_j>0, ~j=1,2,...J,$ being simple zeros of $a(\lambda,0)$.
These proper eigenvalues correspond to $k_j=i\kappa_j, \kappa_j>c.$
The solution of the KdV eq.~(\ref{E:KdV}) is recovered from 
\begin{equation}
\label{soln1}
u(x,t)=c^{2}+2\frac{d}{dx}G(x,x;t) .
\end{equation}
 Using the second linear compatible eq.~(\ref{E:time evolution}) we find the  evolution of the scattering data to be
\begin{align}
 \rho(\lambda; t)&=\rho(\lambda; 0)  e^{i\chi(\lambda)t},
 ~ c_{j}(t)= c_{j}(0)e^{i\chi(\lambda_j)t} ,
\end{align}
where $\chi(\lambda)=8(\lambda^2-c^2)\lambda-4c^{2}\lambda$.

When the eigenvalues are proper ($\kappa_j > c$), solitons will move through the rarefaction ramp and the phase shift can be calculated. For $J=1$ and as $t \rightarrow \infty$, while neglecting the contribution from the continuous spectrum, 
the following one soliton solution is obtained from 
eqs.~(\ref{E:GLM+})-(\ref{soln1}) 
\begin{equation}
\label{KdvSol2A}
u(x;t)\sim c^{2}+2\eta_{1}^{2}~ \text{sech}^{2}\left[ \eta_1 \left(x-( 6c^{2}+4\eta_{1}^{2})t - x_{0}^{+}\right)\right],
\end{equation}
 where $x_{0}^{+}= \frac{1}{2\eta_{1}} \log \left[ -\frac{c_1(0)}{2\eta_{1}} \right]$ with $\lambda_1=i \eta_1 = i \sqrt{\kappa_1^2 - c^2}$ defines the phase of the 1-soliton when $t\to \infty$. 
Thus given a soliton at, say, $t=0$ with phase $x_0$, means the phase shift of the 1-soliton passing through a step is 
$\Delta x_0=x_{0}^{+}-x_0$. A soliton passing through $J$ solitons and a step can be calculated by similar methods (see Appendix \ref{appendC}).
The phase shift from $t=-\infty$ to $t=\infty$ is obtained from the GLM equation formulated from $-\infty ~\text{to}~ x$, which we discuss next.

The GLM equation from $-\infty$ to $x$ is obtained from the right scattering problem for $\psi$ in terms of
$\phi, \overline{\phi}$, 
given by
\begin{equation}
\label{rightscattEq}
\psi(x,k)=\frac{\lambda}{k} \left[ a(k)\overline{\phi}(x,k)-\overline{b}(k)\phi(x,k) \right] .
\end{equation}
We  assume  $\phi$ 
has the following triangular form
\begin{equation}
\label{E:integral equation 3}
\phi(x,k)=e^{-i kx}+\int_{-\infty}^{x}\widetilde{G}(x,s;t)e^{-iks}ds.
\end{equation}
Substituting the above representation  into eq.~(\ref{E:linear combination 2}), {taking into account (\ref{E:symmetry}),} 
 then after dividing by
{$\lambda a(k)/k$} and carrying out analogous calculations to those above yields the following GLM equation,
\begin{equation}
\label{E:GLM2}
\widetilde{G}(x,y;t)+\widetilde{\Omega}(x+y;t)+\int_{-\infty}^{x}\widetilde{\Omega}(s+y;t)\widetilde{G}(x,s;t)ds=0,
\end{equation}
for $x>y$ with kernel
\begin{align}
\label{GLM_kernel}
 \widetilde{\Omega}(z;t)&= \frac{1}{2\pi}\int_{-\infty}^{\infty}\rho_{0}(k;t)e^{-ikz}dk \\ \nonumber
& -\sum_{j=1}^{J}\widetilde{c}_{j}e^{-i k_{j}z}+
\frac{1}{2\pi} \int_{0}^{c} \frac{ \kappa e^{z\kappa}}{\lambda |a(i\kappa;t)|^{2}} d\kappa ,
\end{align}
and 
\begin{equation*}
\rho_0(k,t) = -\frac{\overline{b}(k,t)}{a(k,t)} , ~~~~ \widetilde{c}_{j}(t) =-\frac{i \overline{b}(k_{j}; t)}{\partial_k a(k_{j}; t)},
\end{equation*}
where $k_j=i\kappa_j, \kappa_j>c;$ are simple zero's of $a(k,0)$;
i.e. all the eigenvalues $\kappa_j, j  > 1$ are proper. The third term in 
 eq.~(\ref{GLM_kernel}) 
 is different as compared with the kernel given in eq.~(\ref{KernelGLM+}).
Furthermore this term is associated with the branch cut in the $k-$plane and $-c \leq \lambda \leq c$.

With this GLM equation  the solution of KdV eq.~(\ref{E:KdV}) is given by 
\begin{equation}
\label{soln2}
u(x,t)=-2\frac{d}{dx}\widetilde{G}(x,x;t)  .
\end{equation}
The time evolution of the data is given by
\begin{align}
\label{TimeEvolRight1}
 \rho_{0} &(k;t) =\rho_{0}(k,0)e^{-8ik^{3}t},  ~\tilde{c}_{j}(t)=\tilde{c}_{j}(0)e^{-8ik_j^3t} , \\
\label{TimeEvolRight2}
&  |a(i\kappa,t)|=|a(i\kappa,0)|e^{4 \kappa^3 t },  ~~ \text{for} ~~ 0 \le \kappa \le c .
\end{align}
The soliton phase shift of proper solitons as $t\rightarrow -\infty$ can be obtained from the above GLM equation. Assuming no pseudo-embedded eigenvalues, and neglecting  the contribution from the continuous spectrum, 
the following one soliton solution is obtained from eqs.~(\ref{E:GLM2})-(\ref{soln2})
\begin{equation}
\label{KdvSol2B}
u(x;t)\sim 2\kappa_{1}^{2}~\text{sech}^{2}\left[\kappa_{1}(x-4\kappa_{1}^{2}t-x_{0}^{-})\right],
\end{equation}
where 
$x_{0}^{-}=  \frac{1}{2\kappa_{1}} \log\left[ - \frac{2\kappa_{1}}{\tilde{c}_1(0)} \right]$.
Hence the total phase shift from $-\infty$ to $\infty$ across the step is given by $\Delta x_0=x_{0}^{+}-x_{0}^{-}$ where $x_{0}^{+}$  is given below equation (\ref{KdvSol2A}). It should be noted that  this formula contains the step and its associated continuous spectra contributions which are encoded into the normalization constants $c_j(0),\tilde{c}_j(0)$; we also note that the above phase shift formulae agree with numerical simulations \cite{MJAJCXL2017}.

\section{trapped solitons and pseudo-embedded eigenvalues} \label{pseudo_emebed_sec}
In this section we show how an initial soliton/sech$^2$ profile with amplitude parameter  $0<\kappa_0<c$ ~can be described using the GLM approach. 
To this end we consider the GLM equation from $-\infty ~\text{to}~ x$ given in  (\ref{E:GLM2}). 
We assume no proper eigenvalues; so we only have two terms, both from the continuous spectrum: the first and last terms in 
(\ref{GLM_kernel}). The last term arises due to the branch cut. 

We assume a localized initial condition: a soliton/sech$^2$  form with corresponding 
amplitude parameter $\kappa_0$, or a box function or a delta function, located well to the left of the step centered at $x=0$; we call $k_0 = i\kappa_0$ a pseudo-embedded eigenvalue. Additional solitons/${\rm sech}^2$ profiles can be added in a similar manner. 
The first term in the kernel (\ref{GLM_kernel}) is small in the neighborhood of this pulse, so we only need to consider the branch cut contribution, i.e. the third term. 

The dominant contribution to this integral,  comes from values of $\kappa$ near $\kappa_0$ where $a(i \kappa)$ is nearly zero. 
We substitute the form of 
$a(k)$ given in eq. (\ref{a(k)}) (see Appendix \ref{appendA} for more details) into the branch cut integral in (\ref{GLM_kernel}), expand around the point $k = i \kappa_0$, 
and insert the time dependence of $a(k,t)$ from equation (\ref{TimeEvolRight2}) into this term, which we call $ \widetilde{\Omega}_3(z;t).$ The dominant contribution is given by  the integral
\begin{equation}
\label{BrCut}
\widetilde{\Omega}_3(z;t) \sim \frac{\kappa_0 e^{\kappa_0 z-8\kappa_0^3t}}{2\pi \epsilon \lambda_0} \int_{-\infty}^{\infty} \frac{ 1} { \Delta(\kappa')}d\kappa' \; , 
\end{equation}
where $\lambda_0 = \sqrt{c^2 - \kappa_0^2}$, 
\begin{equation}
\label{Delta}
\Delta (\kappa') = \\|a_1|_0^2 \kappa'^2 - 2|a_1|_0|a_2|_0\sin(\varphi_1-\varphi_2)\kappa' +\\ |a_2|_0^2,
\end{equation}
 and  $\kappa-\kappa_0= \epsilon \kappa', ~a_j(i\kappa_0)=|a_j|_0e^{i\varphi_j},~j=1,2.$ We note that $|a(i\kappa,0)|^2$ given in the third term of the kernel  (\ref{GLM_kernel}) is approximated by $\Delta (\kappa)$  in the neighborhood of $\kappa_0$. 
Evaluating the above integral we find 
\begin{equation}
\label{BrCutResult}
\widetilde{\Omega}_3(z;t)\sim \frac{\kappa_0}{2\lambda_0\alpha \epsilon}e^{z \kappa_0-8\kappa_0^3t}.
\end{equation}
where  $\alpha=|a_1|_0|a_2|_0|\cos(\varphi_1-\varphi_2)|>0.$
Remarkably, this has {\it exactly the same form} as that from the discrete spectra in the GLM equation given in (\ref{GLM_kernel}). 
The corresponding solution is given by 
\begin{equation}
\label{KdvSol2B}
u(x,t)\sim 2\kappa_{0}^{2}~ \text{sech}^{2}\left[\kappa_{0}(x-4\kappa_{0}^{2}t - x_{0}^{-})\right],
\end{equation}
where 
\begin{equation}
\label{PhSftx0-}
x_0^{-}=  \frac{\ln\left( 4 \lambda_0 \alpha \epsilon \right)}{2 \kappa_0 }  ,
\end{equation}
and valid when $x_0 \ll -1$ and the soliton position $x-x_0$ is well to the left of the ramp.

Thus, far to the left of the step ($x_0 \ll -1$), a soliton-like pulse travels with pseudo-eigenvalue $\kappa_0$.
This soliton/${\rm sech}^2$ mode 
travels unimpeded until it comes into contact with the ramp that emanates from the step up initial condition. This soliton/${\rm sech}^2$ becomes trapped by the ramp (see Fig.~\ref{trap_soliton}).
We refer to this as a trapped soliton. To carry out the details of this long time asymptotic analysis of the trapping from the inverse problem is outside the scope of this paper. The weakly dispersive case is discussed in \cite{MJAJCXL2017} where analysis and numerical calculations further show how the soliton becomes trapped in the ramp and never never makes it to the top of the ramp.

\subsection{Conclusion}
The scattering/inverse scattering theory associated with the time-independent Schr\"odinger equation and its relationship to soliton solutions of the KdV equation for step potentials was analyzed.
 
The first case we considered was that of ``proper'' eigenvalues: 
$a(k)=0, k=i\kappa; \kappa>c$ where we find ``proper'' solitons. 
 In this case the inverse scattering theory and linearization of the KdV equation can be carried out via a GLM equation with the solitons calculated from the discrete spectrum of the GLM kernel. Here a soliton that is initially well separated from the step propagates all the way through a ramp; 
doing so it acquires a phase shift which can be calculated exactly. This phase shift has encoded in it the continuous spectra which arises from the step. 
 Numerical calculations confirm these formulae \cite{MJAJCXL2017}. 

The second case was that of spectral data which had no proper eigenvalues, yet behaved as though it did. In terms of the soliton pulse, the amplitude 
is not large enough to pass though the rarefaction ramp that develops from the step up initial condition. This becomes a trapped soliton. In spectral terms there is a point, $k=i\kappa, 0 \leq \kappa \leq c$, where the inverse of the transmission coefficient: $a(k),$ is exponentially close to, but {\it not} zero. 
In this case the continuous spectrum associated with the branch cut $0 \leq \Im k \leq c$ gives rise to a 
contribution  that approximates 
a discrete eigenvalue located at $k=i\kappa_0$. We call such $\kappa_0$ a pseudo-embedded eigenvalue.

 Although the analysis here is developed for step up boundary conditions, the step down case  is similar.  In \cite{Hoefer2016} the correspondence between the step up and step down case is referred to as `hydrodynamic reciprocity'.
 In the step down case a localized initial 
 profile is inserted to the right of the step. Upon evolution it gets trapped by a DSW that emanates out of the initial data. 
 From a mathematical viewpoint, we have the relationship  $\lambda^2=k^2-c^2$ for step down boundary data (compare this with eq.~(\ref{lambda})). Here the scattering/inverse scattering theory corresponds to interchanging the roles of $\lambda$ and $k$.

\section{Acknowledgements}
MJA was partially supported by NSF under Grant No. DMS-1712793.

\bibliographystyle{amsplain}

\appendix

\section{Examples of scattering data}
\label{appendA}
In this appendix we discuss three examples of potentials associated with the time-independent Schr\"odinger equation (\ref{E:Schrod}). These potentials are given as initial conditions associated with the KdV equation  (\ref{E:KdV}) at $t=0$.
Each consists of a localized hump well separated from a step with the  form
\begin{equation}
u(x,0)= u_0(x,x_0)+c^2H(x) ,
\end{equation}
where the Heaviside function $H(x)$ is given by
\begin{equation}
\label{H(x)}
H(x) = 
\begin{cases}
0 , & x<0  \\
1 , & x>0  \\
\end{cases} \; .
\end{equation}
Here $x_0$ represents the center of the localized hump and is assumed to be located far to the left of $x = 0.$ 

\subsection{Soliton/sech$^2$ potential}
Consider a soliton/sech$^2$ profile positioned well to the left of the step function: 
\begin{equation}
\label{SolPlusStepFn}
u(x)= 2\kappa_0^2~ {\rm sech}^2 \left[ \kappa_0(x-x_0) \right] +c^2H(x), 
\end{equation}
where $-x_0 \gg 1, c>0,\kappa_0  > 0 $.

If we consider the decaying problem ($c = 0$) 
corresponding  to the above sech$^2$ potential we can calculate $\phi(x,k), \psi(x,k)$ exactly. The reason for this is that when $b(k)=0$ in eq.~(\ref{E:linear combination 1}), or when $\overline{b}(k)=0$ in eq.~(\ref{rightscattEq}), there are significant simplifications.  In eqs.~(\ref{E:linear combination 1}),  (\ref{rightscattEq}) we divide by $a(k)$, subtract the pole contributions, use the symmetries in  (\ref{E:symmetry}) and take a minus projector.  Evaluation at $k=i\kappa_0$ yields the bound state and then using the bound state for general $k$ we can calculate the eigenfunction (cf. \cite{MJA2011}). In this way we find the eigenfunction $\phi(x,k)$ which is valid for $x \leq 0$ in the step problem; similarly we can get $\psi(x,k)$.  However for $\psi(x,k)$ in the step problem $k$ is replaced by $\lambda$ since the spectral parameter satisfies 
 $\lambda^2=k^2+c^2$. The results are
 \begin{equation}
 \label{phidecay}
 \phi(x,k)=e^{-ikx}\left( 1-  \frac{2i\kappa_0}{k+ i\kappa_0}  \frac{1}{1+e^{-2\kappa_0(x-x_0)} } \right), ~  x \leq 0                 
  \end{equation}
\begin{equation}
 \label{psidecay}
 \psi(x,\lambda)=e^{i\lambda x}\left( 1- \frac{2i\kappa_0}{\lambda+ i\kappa_0}  \frac{1}{1+e^{2\kappa_0(x-x_0)} }\right),  ~  x \geq 0  .       
\end{equation}
From these results we can calculate the scattering data using the Wronskian. Using eq.~(\ref{E: scattering data}) $a(k)$ is found to be
 \begin{equation}
 \label{a(k)sol}
 a(k)=\frac{c^2k+(k+\lambda)(k^2+ \kappa_0^2)+ic^2\kappa_0 \rm{tanh}(\kappa_0 x_0)}{2\lambda (k+i\kappa_0)(\lambda+i\kappa_0)} .
  \end{equation}
From the above relation we can calculate the first two terms of the  asymptotic approximation of the form given by eq.~(\ref{a(k)}) which for convenience  we give again below
\begin{equation*}
a(k)= a_1(k)(k-i\kappa_0)+\epsilon a_2(k), ~ \epsilon=e^{2\kappa_0 x_0} \ll 1 .
\end{equation*}
The values $a_1(k),a_2(k)$ for the soliton plus step (\ref{SolPlusStepFn}) are given by
 \begin{equation}
  \label{a1,a2sol}
  a_1(k)= \frac{\lambda+k}{2\lambda(k+i\kappa_0)}, ~a_2(k)=\frac{i c^2 \kappa_0}{\lambda(k+i\kappa_0)(\lambda+i\kappa_0)} .
  \end{equation}
  
To approximate the branch cut integral in (\ref{GLM_kernel}) in the case of pseudo-embedded eigenvalues we focus on values of $k$ near $ i \kappa_0$ since that is where $a(k)$ is at a minimum. As such, to get the asymptotic integral in eq.~(\ref{BrCut}) we expand $a(k)$ around $k = i \kappa_0$  which results in evaluating $a_1(k)$ and $a_2(k)$ in eq.~(\ref{a(k)}) at $k = i \kappa_0$.

 We remark that a similar example can also be calculated exactly, namely that of a soliton truncated at zero at $x=0$: 
  \begin{equation}
\label{SolPlusStepFn}
u(x)= 2\kappa_0^2~ {\rm sech}^2 \left[ \kappa_0 (x-x_0) \right][1 - H(x) ]+c^2H(x) .
\end{equation}
In this case $\phi(x,k)$ is still given by eq.~(\ref{phidecay}) and $\psi(x,\lambda)=e^{i\lambda x}$ for $ x\geq 0$. Hence $a(k)$ can be calculated from the Wronskian relation. The formula is 
\begin{equation}
a(k) = \frac{ 2 \kappa_0^2 +(k + \lambda) \left[k + k \cosh(2 \kappa_0 x_0) + i \kappa_0 \sinh(2 \kappa_0 x_0) \right] }{ 4 \lambda (k + i \kappa_0) \cosh^2(\kappa_0 x_0)} ,
\end{equation}
which has the same $ a_1(k)$ as (\ref{a1,a2sol}), but different $a_2(k)$.

We also note that while there are solutions 
$a(k)=0$ for $\Im k > c$ we do not find any solutions to $a(k)=0$ when $0 < \Im k <c$. 
An asymptotic expansion suggests that the zeros of $a(k)$ are complex i.e. $k_0 = \xi_0 + i \kappa_0$ where $\kappa_0 >0 , \xi_0 \not=0$.
This is, in fact, a contradiction since any eigenvalue corresponding to a bound state must be purely imaginary (see Appendix \ref{appendB}).

\subsection{Delta potential}
Consider a delta function of height $Q$ positioned well to the left ($  -x_0  \gg 1$) of a step function: 
\begin{equation}
\label{DeltaFn}
u(x)= Q\delta(x-x_0)+c^2H(x), ~~~ Q > 0 .
\end{equation}

The time-independent Schr\"odinger equation can be explicitly calculated in this case. The solution takes the form
\begin{equation}
\label{Delta_soln}
\phi(x,k) = 
\begin{cases}
e^{-ikx}, &  x < x_0 \\
\alpha_1(k) e^{- i k x} + \beta_1(k) e^{  i k x}, &  x_0<x<0 \\
a(k) e^{- i \lambda x} + b(k) e^{  i \lambda x} , & x > x_0  \\
\end{cases} \; ,
\end{equation}
for $\lambda = \sqrt{c^2 + k^2}.$ At $x=x_0$ the eigenfunction $\phi(x,k)$ satisfies the jump condition
$[\partial_x \phi(x,k)]_{  x_0^-}^{x_0^+}+Qe^{ - i k x_0}=0$ and continuity. Using continuity of $\phi(x,k)$ and its derivative at $x=0$ yields the remaining coefficients. We only give  $a(k)$ for the delta function plus step below
\begin{equation}
\label{Delta_soln_a}
a(k)=  \frac{\lambda+k}{2 \lambda}\left( 1+\frac{Q}{2ik} \right) + \frac{\lambda-k}{2 \lambda} \left( -\frac{Q}{2ik}e^{-2ikx_0} \right) .
\end{equation}

In the case of proper eigenvalues there exists $k_1 = i \kappa_1 , \kappa_1 > c$ such that $a(k_1) = 0.$ For $-x_0 \gg 1, \kappa_1 \approx  Q/2$ which is the same as the decaying (non-step) problem. 
When $0<\kappa_0<c$ the unperturbed  problem,  i.e. without  the exponential term in equation (\ref{Delta_soln_a}), 
suggests that $\kappa_0$  should be approximated by $Q/2$. But
keeping the exponential term and carrying out an asymptotic expansion for $-x_0 \gg 1$ leads to the zeros of $a(k)$ being complex i.e. $k_0 = \xi_0 + i \kappa_0$ where $\kappa_0 >0$ {\it and} $\xi_0 \not=0$.
This is  a contradiction since any true bound state eigenvalue is purely imaginary (see Appendix \ref{appendB}). 
In fact trying  to solve $a(k)=0$ numerically does not lead to a convergent iteration 
for this or any of the pseudo-embedded/trapped soliton examples discussed in this appendix.

From formula (\ref{Delta_soln_a})  we can calculate the pseudo-eigenvalue and the first two terms of the  asymptotic approximation given in eq.~(\ref{a(k)}); they are
\begin{equation}
 \label{a1,a2delta}
 \kappa_0=\frac{Q}{2},~ a_1(k)= \frac{\lambda+k}{2\lambda k}, ~a_2(k)=\frac{i\kappa_0(\lambda-k)}{2\lambda k} .
 \end{equation}

\subsection{Box potential}
Consider a box function positioned well to the left ($x_0 \ll -1$) of a Heaviside function (\ref{H(x)}):  
\begin{equation}
\label{BoxFn}
u(x)= h^2 B(x - x_0)+c^2H(x), 
\end{equation}
 \begin{equation*}
\label{B(x)}
B(x - x_0) = 
\begin{cases}
0 , & |x - x_0  | > L/2  \\
1 , & |x - x_0  | \le L/2  
\end{cases} ,
\end{equation*} 
with height $h^2, h>0$ and width $L>0$.
The solution takes the form
\begin{equation}
\label{Box_soln}
\phi(x,k) = 
\begin{cases}
e^{-ikx}, &  x < x_0 - L/2 \\
\alpha_1(k) e^{- i \eta x} + \beta_1(k) e^{  i \eta x} , &  | x - x_0| < L/2 \\
\alpha_2(k) e^{- i k x} + \beta_2(k) e^{  i k x}  ,& x_0 + L/2 < x < 0 \\
a(k) e^{- i \lambda x} + b(k) e^{  i \lambda x} , &   x >0
\end{cases} \; ,
\end{equation}
where $\eta=\sqrt{h^2+k^2}$. We enforce continuity of the solution and its derivative at $x=x_0$ and $x=0$. All coefficients can be calculated.
We only give  $a(k)$ for this case below
\begin{align}  \nonumber
 a(k) = - & \frac{(\lambda + k)}{8 \lambda \eta k} e^{i k L} \left[ (\eta - k)^2 e^{i \eta L} - (\eta +k)^2 e^{- i \eta L}  \right] \\
 \label{Box_soln_a}
 +& \frac{(\eta^2 - k^2) (\lambda - k) }{8 \lambda \eta k} \left[ e^{i \eta L} - e^{- i \eta L} \right] e^{- 2 i k x_0} .
\end{align}

If we neglect the term multiplied by $ e^{- 2 i k x_0}$ i.e. take $x_0 \ll -1$, the eigenvalues satisfying $a(k)=0$ yield solutions for $\eta \in \mathbb{R}$ satisfying 
\begin{equation}
\label{boxevs} 
\tan(\eta L) = - \frac{2 i k \eta}{k^2 + \eta^2} .
\end{equation}
Solutions to the above equation are the same 
as those obtained in the decaying no-step problem \cite{MJA2011}. Graphical analysis shows that there can be one or more solutions depending on the size of $h$ and $L$. For example, when $h<\pi/L$ there is one solution, which we denote as $\kappa_0$ for $0 < \kappa_0 < h$.

Keeping the exponential term modifies the above result. For $\eta \in \mathbb{R}, k=i \kappa, \kappa>c$ define $\lambda=i \tilde{\lambda}, \tilde{\lambda} \in \mathbb{R}$. Solutions of $a(k)=0$ must satisfy 
\begin{equation}
\label{boxevs2} 
\left( \frac{\eta+i\kappa}{\eta-i\kappa}\right)^2 \left(\frac{1- \frac{(\eta^2+\kappa^2)(\tilde{\lambda}-\kappa)}{(\eta+i\kappa)^2(\tilde{\lambda}+\kappa) }        e^{2\kappa x_0 + \kappa L}     }{1-\frac{(\eta^2+\kappa^2)(\tilde{\lambda}-\kappa)} {(\eta-i\kappa)^2(\tilde{\lambda}+\kappa)}  e^{2\kappa x_0 + \kappa L }  }\right)   =e^{2i\eta L} .
\end{equation}
Note that both left and right sides of the above formula have unit modulus and hence a perturbative solution for $-x_0 \gg 1$
is expected; numerical solutions have been found. 

For $\eta \in \mathbb{R}, k=i \kappa, 0<\kappa<c, \lambda \in \mathbb{R}$ solutions of $a(k)=0$ now satisfy 
\begin{equation}
\label{boxevs3} 
\left(\frac{\eta+i\kappa}{\eta-i\kappa}\right)^2 \left(\frac{1-\frac{(\eta^2+\kappa^2)(\lambda-i\kappa)}{(\eta+i\kappa)^2(\lambda+i\kappa)}      e^{2\kappa x_0 + \kappa L}        }{1- \frac{(\eta^2+\kappa^2)(\lambda-i \kappa)} {(\eta-i\kappa)^2(\lambda+i\kappa)} e^{2\kappa x_0 + \kappa L}    }\right)   =e^{2i\eta L} .
\end{equation}
In this case the left-hand side is not of unit magnitude; no solution is expected; a numerical 
solution has not been found. The values of $a_1(k)$ and $a_2(k)$ can be found from equation (\ref{Box_soln_a}).
 
\section{Bound states}
\label{appendB}

In this appendix we establish that all solutions of $a(k)=0$ associated with bound states, i.e. eigenvalues, must be purely imaginary with $\Im k>0$.
Recall that
\begin{align*}
 &\phi(x, k) \sim e^{-ik x}  , ~~~ x\to -\infty, \\
 & \psi (x, \lambda)\sim e^{i \lambda x}  , ~~~ x\to +\infty ,
\end{align*}
and $ a(k)=\frac{1}{2i\lambda}W(\phi, \psi)$ for the Wronskian $W(\phi, \psi) = \phi \psi_x-\phi_x \psi.$ 
If $a(k_{0})=0$ and $k_{0}=\xi_{0}+i\kappa_{0}$ for $\xi_0 \not = 0$, then $\phi (x, k_{0})=\beta_{0}\psi(x, k_{0})$ for some nonzero constant $\beta_{0}$. Thus,
\begin{equation}
\phi(x, k_{0}) \sim e^{-i \xi_{0}x} \cdot e^{\kappa_0 x}\ \ \ \text{as} \ \ \ x\to -\infty,
\end{equation}
and
\begin{equation}
\phi(x, k_{0}) \sim \beta_{0}e^{i \lambda_{0}x}=\beta_{0} e^{-\Im \lambda_{0} x}\cdot e^{i \Re \lambda_{0} x} \ \ \ \text{as} \ \ \ x\to +\infty,
\end{equation}
where $\lambda_{0}:=\sqrt{k_{0}^{2}+c^{2}}$. Assuming that 
\[ \phi(x, k_{0}) \to 0, ~~~  |x| \to \infty , \]
then necessarily $\Im \lambda_{0}>0$ and hence $\kappa_{0}>c$. Such eigenvalues are proper. Moreover, we point out that $\phi(x,k)$ and its complex conjugate $\phi^{*}(x, k)$ satisfy the equations 
\begin{align*}
& \phi_{xx}+\left(u(x)+ k^{2} \right)\phi=0, \\
& \phi_{xx}^{*}+\left(u(x)+ (k^*)^{2} \right)\phi^{*}=0,
\end{align*}
respectively, with $u(x)$ real. Hence, 
\begin{equation}
\frac{\partial}{\partial x}W(\phi, \phi^{*})+\left( (k^*)^2-k^{2} \right)\phi\phi^{*}=0.
\end{equation}
Since $\phi\to 0$, $\phi_{x}\to 0$ as $x \to \pm \infty$, we have 
\begin{equation}
\left((k^*)^{2}-k^{2} \right)\int_{-\infty}^{\infty}|\phi(x,k)|^{2}dx=0.
\end{equation}
If $a(k_{0})=0$, where $k_{0}=\xi_{0}+i\kappa_{0}$, then 
\[\xi_{0}\kappa_{0}\int_{-\infty}^{\infty}|\phi(x,k)|^{2}dx=0 , \] 
for $\phi(x,k) \in L^2(\mathbb{R})$ so $\int_{-\infty}^{\infty}|\phi(x,k)|^{2}dx>0$, and $\xi_{0}\kappa_{0}=0$.  For decay as $|x| \to \infty$ we require $\kappa_{0}>0$; thus $\xi_{0}=0.$

\section{Phase shift for $J$ solitons}
\label{appendC}

\subsection{J-soliton solution from the GLM equation (\ref{E:GLM+})}
We take $J$ ordered proper eigenvalues, i.e. $J$ distinct 
simple zeros of $a(\lambda)$.
We assume that each soliton is  initially well separated and located well to the left 
of the step.
Neglecting all 
effects due to the continuous spectrum, 
the kernel $\Omega$ in 
the GLM equation (\ref{E:GLM+}) is 
\begin{equation}
\Omega(z;t)=-\sum_{j=1}^{J} c_{j}(0)e^{(8\eta_{j}^{3}+12c^{2})t - \eta_j z},
\end{equation}
where $\lambda_{j}=i\eta_{j}$, $\eta_{j}>0$ and $\eta_1<\eta_2<....<\eta_J$, and $c_j(0)$ is defined below eq.~(\ref{KernelGLM+}). Then as $t\to\infty, x\sim 4\kappa_J^2t$
we find that the fastest, or $J$th soliton, is asymptotically given by \cite{AblKod82}

\begin{equation}
u_{J}(x;t)\sim c^{2}+2\eta_{J}^{2}~ \text{sech}^{2} \left[\eta_{J}\left(x-(6c^{2}+4\eta_{J}^{2})t-x_{J}^{+}\right)\right],
\end{equation}
where 
\begin{equation}
\label{ps+}
\eta_J x_{J}^{+} = \frac{1}{2}\log \left[ - \frac{c_J(0)}{2\eta_{J}} \right] 
+ \sum_{j=1}^{J-1}\log\left|\frac{\eta_{J}-\eta_{j}}{\eta_{J}+\eta_{j}}\right| ,
\end{equation}
defines the phase of $J$th soliton when $t\to +\infty$.



\subsection{J-soliton solution from the GLM equation (\ref{E:GLM2})}
As above,  we take $J$ eigenvalues, i.e. $J$ distinct simple zeros of $a(k)$: $a(k_j=i\kappa_j)=0$. 
We assume that each soliton is initially well separated from every other soliton 
and neglect the effects due to the  the continuous spectrum: the first and third terms in the kernel $\Omega$ given in equation (\ref{GLM_kernel}). We consider $J$ ordered solitons that are initially separated well to the left of the step. The kernel is given by
\begin{equation}
\Omega(z;t)=-\sum_{j=1}^{J} \tilde{c}_{j}(0)e^{\kappa_j z- 8\kappa_{j}^{3}t} , 
\end{equation}
where 
$\kappa_{j}>0$ and $\kappa_1<\kappa_2<....<\kappa_J$ where $\tilde{c}_j(0)$ is defined below (\ref{GLM_kernel}). 
As $t\rightarrow -\infty, x\sim 4\kappa_J^2 t$ the $J$th soliton satisfies
\begin{equation}
u_{J}(x;t)\sim 2 \kappa_{J}^{2} ~ \text{sech}^{2}\left[\kappa_{J}(x-4\kappa_{J}^{2}t-x_{J}^{-})\right],
\end{equation}
where 
\begin{equation}
\label{ps-}
 \kappa_Jx_{J}^{-}  =\frac{1}{2}\log\left[ - \frac{2\kappa_{J}}{\widetilde{c}_{J}(0)} \right] 
-  \sum_{j=1}^{J-1}\log\left|\frac{\kappa_{J}-\kappa_{j}}{\kappa_{J}+\kappa_{j}}\right| ,
\end{equation}
defines the phase of $J$th soliton when $t\to -\infty$. 
Thus, the total phase shift of $J$th soliton due to the step and other solitons is given by
\begin{equation}
\label{phseshftJsol}
\Delta x_{J} = \eta_{J}x_{J}^{+}-\kappa_{J}x_{J}^{-} .
\end{equation}

\end{document}